\documentclass[12pt]{article}
\usepackage{latexsym,graphicx,color}
\usepackage{amsmath, amssymb, bm,eucal}
\begin{document}
\begin{flushright}
USTC-ICTS-14-18\\
\end{flushright}
\vspace{20mm}

\begin{center}

{\Large \bf Baryons in the Sakai-Sugimoto model in the D0-D4 background}

\vspace{10mm}

{\large Wenhe Cai$^\dagger$\footnote{Email:edlov@mail.ustc.edu.cn}, Chao
Wu$^\dagger$\footnote{Email:wuchao86@mail.ustc.edu.cn}, and Zhiguang
Xiao$^{\dagger,\ddagger}$\footnote{Email:xiaozg@ustc.edu.cn} 
\\[4mm]
$^\dagger$Interdisciplinary Center for Theoretical Study,\\
 University of Science and Technology of China,
\\ Hefei, Anhui
 230026, China
\\[2mm]
$^\ddagger$State Key Laboratory of Theoretical Physics,
\\
 Institute of Theoretical Physics, Chinese Academy of Sciences,
\\Beijing, 100190, China
}

\end{center}

\vspace{10mm}

%\title{Baryons in Sakai-Sugimoto model in D0-D4 background}
%\author{Wenhe Cai$^1$\footnote{Email:edlov@mail.ustc.edu.cn}, Chao
%Wu$^1$\footnote{Email:wuchao86@mail.ustc.edu.cn}, and Zhiguang
%Xiao$^{1,2}$\footnote{Email:xiaozg@ustc.edu.cn} 
%\\[4mm]
%$^1$Interdisciplinary Center for Theoretical Study,\\
% University of Science and Technology of China,
%\\ Hefei, Anhui
% 230026, China
%\\[2mm]
%$^2$State Key Laboratory of Theoretical Physics,
%\\
% Institute of Theoretical Physics, Chinese Academy of Sciences 
%}
%\maketitle
%
\abstract{
The baryon spectrum in the Sakai-Sugimoto model in the D4 background
with smeared D0 charges is studied. We follow the instanton description of
baryons by Hata et al.[Prog. Theor. Phys. 117, 1157].  The background corresponds to an
excited state with nonzero glue condensate $\langle {\rm tr} (F_{\mu\nu}\tilde
F^{\mu\nu})\rangle$ which is proportional to the D0 charge density. The
baryon size shrinks when we turn on small D0 charge density.
But for larger D0 charge density where massive modes in the gauge
theory may also take effect, the size of baryons will grow. The difference
between baryon masses will become smaller when D0 charge density
increases. There may also be indications that the baryon will become
unstable and cannot exist for sufficiently large D0 density. 
}

\section{Introduction}
In recent years, with the running of the RHIC, there have been some
discussions on the spontaneous parity violation in hot QCD.  
Some proposals were put forward that $P$- or $CP$- odd bubble 
may be created during the collisions
\cite{Kharzeev:1998kz,Buckley:1999mv,Kharzeev:2004ey}. A metastable
state with nonzero QCD vacuum $\theta$ angle or $tr (F_{\mu\nu}\tilde F^{\mu\nu})$ could
be produced in some space-time region in the hot and dense condition
when deconfinement happens. According to the proposal, with the rapid
expansion of the bubble, it cools down and the metastable state
freezes inside the bubble\cite{Buckley:1999mv,Shuryak:2001jh},
returning to the confinement phase. Then a $P$- or $CP$-odd bubble
forms and may soon decay into the true vacuum. Chiral
magnetic effect was proposed as a test of this kind of phenomenon
\cite{Kharzeev:2007jp,Fukushima:2008xe}. See
\cite{Kharzeev:2013ffa} for a review.

The state with nonzero  $tr (F_{\mu\nu}\tilde F^{\mu\nu})$ may also
play a role in the confinement mechanism. Many
mechanisms were proposed as the possible cause of the confinement. See
\cite{Simonov:1997js} for a review. Among these mechanisms, some 
classical or semiclassical gauge field configurations are 
important, such as some topologically nontrivial
solutions---monopoles, instantons, and so on. In particular, there could be
field configurations with constant field strength as solutions for the classical
equation of motion or at the minimum of effective potential.  Self-dual field strength is studied in
\cite{Leutwyler:1980ev,Minkowski:1981ma,Flory:1983dx,vanBaal:1984ar},
and was proposed to be a mechanism for the confinement
\cite{Efimov:1998hi}. Some of these solutions come with
nonzero $tr
(F_{\mu\nu}\tilde F^{\mu\nu})$. There can also be domain
solutions\cite{Leutwyler:1980ev,Ambjorn:1979xi,Amundsen:1990nt}
that have different field strength pointing in diverse directions in
different tiny space-time domains. %and on average, 
By averaging over a larger length scale, some global symmetries, such as Poincar\'e symmetry,
 are recovered. In our paper, we are interested in the
states with only $P$-parity non-conserved and the Poicar\'e invariance
 preserved, that is, states with nonzero $\langle tr (F_{\mu\nu}\tilde
F^{\mu\nu})\rangle$ and vanishing $\langle F_{\mu\nu}\rangle$.

To study the effects of the states with
nonzero $tr (F_{\mu\nu}\tilde F^{\mu\nu})$, one has to resort to some 
nonperturbative methods. String-gauge duality
provides a powerful nonperturbative method to study this kind of phenomena. Adding $\langle tr
(F_{\mu\nu}\tilde F^{\mu\nu})\rangle $ condensate to gauge theory was
first studied in N=4 SUSY YM along these lines,
and corresponds to adding smeared D(-1) charges into D3-brane
configuration \cite{Liu:1999fc,Kehagias:1999iy}. Temperature and flavors can also be introduced into the
Yang-Mills, and then quark condensates, meson spectra,  baryon
properties, etc., could be
studied\cite{Kehagias:1999iy,Ghoroku:2004sp,Ghoroku:2005tf,Brevik:2005fs,Ghoroku:2006af,Erdmenger:2007vj,Erdmenger:2011sz,Ghoroku:2008tg,Ghoroku:2008na,Sin:2009yu,Gwak:2012ht,Sin:2009dk}.

Several other holographic constructions of the QCD-like theory are based on the D4 background initiated by
Witten \cite{Witten:1998zw}, which corresponds to the five-dimensional gauge theory compactified on
a small circle to give a four-dimensional gauge theory. Among others,
Sakai and Sugimoto
proposed a promising model, which realizes the spontaneous chiral symmetry
breaking by a geometric construction \cite{Sakai:2004cn,Sakai:2005yt}.  
In this model,  flavors
 are realized by introducing $N_f$ D8-branes and $N_f$ anti-D8-branes
which combine at the tip of the cigar geometry of the D4 soliton.  
This geometry is supposed to describe the spontaneously breaking of
the $U_{L}(N_f)\times U_{R}(N_f)$ symmetry
to $U_V(N_f)$ in the four-dimensional theory, which is
strengthened by the existence of the massless Goldstones in the
spectrum. The low energy chiral Lagrangian for mesons can also be derived in
this model. Similar to
the D(-1)-D3 background, adding condensate $\langle tr
(F_{\mu\nu}\tilde F^{\mu\nu})\rangle$ in the S-S model corresponds to
adding smeared D0 charges into the D4 background. The gauge theory
in this background is studied in \cite{Barbon:1999zp,Suzuki:2000sv}.
Putting Sakai-Sugimoto model (S-S model) into this background  allows
us to study the hadron phenomena in the nonzero $\langle tr
(F_{\mu\nu}\tilde F^{\mu\nu})\rangle$ background.  We have already
studied the meson spectra and the interactions of the lowest-lying
vector mesons and Goldstones in this background in \cite{Wu:2013zxa}
and found out that introducing $\langle tr (F_{\mu\nu}\tilde F^{\mu\nu})\rangle$ does not
change the property of the spontaneous breaking of the $SU_L(N_f)\times
SU_R(N_f)\to SU_V(N_f)$. In
the present paper, we will study the baryon mass spectrum in this background.
As in \cite{Wu:2013zxa},
to keep the $\langle tr (F_{\mu\nu}\tilde F^{\mu\nu})\rangle$
dependence in the large $N_c$, we require it to be of ${\mathcal
O}(N_c)$ as in \cite{Liu:1999fc}, $\tilde \kappa\sim\langle tr
(F_{\mu\nu}\tilde F^{\mu\nu})\rangle/ N_c $. Baryons
can be turned on in the S-S model by  introducing the baryonic D4-branes wrapping the sphere
directions, which  can be related to the Skyrmions in the low energy
effective theory\cite{Sakai:2004cn,Nawa:2006gv}.
This can be realized as the soliton solutions of the gauge fields
on D8\cite{Hata:2007mb},  and the nucleon
interactions can also be
modelled along these lines\cite{Hashimoto:2008zw,Hashimoto:2009ys,Kaplunovsky:2010eh}.
In 
\cite{Seki:2013nta}, uniform distributed baryon system in the D0-D4 
background is studied and
the chiral condensate is discussed.  
We will follow the approach of \cite{Hata:2007mb}, and study the
$\tilde \kappa$ dependence of the baryon mass spectrum. We will see
that there may be indications that the baryons may not be able to
exist with strong $\tilde \kappa$ turned on with massive KK modes in
the gauge theory included.

The structure of this paper is as follows: In section \ref{sect:D0D4}, we
review the D0-D4 background and its relation to the gauge field
theory following \cite{Wu:2013zxa}. In section \ref{sect:classical},
we put the S-S model in this background and we study the soliton
solution of the action at order of $\lambda^0$. In section
\ref{sect:quantum} we use the collective coordinate quantization to
obtain the mass spectrum of the baryons and their $\tilde
\kappa$ dependence. Section \ref{sect:discussion} is the
conclusion.

\section{The D0-D4 background and the corresponding field theory
\label{sect:D0D4}}
In this section,  we review the D0-D4 background and its corresponding
field theory following \cite{Barbon:1999zp}. 

The solution of  D4-branes with smeared D0 charges in type IIA
supergravity in Einstein frame is \cite{Barbon:1999zp,Suzuki:2000sv}
\begin{eqnarray}
ds^2&=& H_4^{-\frac3 8}\left(- H_0^{-\frac 7 8} f(U) d\tau^2+H_0^{\frac 1
8}\Big((dx^0)^2+(dx^1)^2+\cdots+(dx^3)^2\Big)\right)
\nonumber\\&&
+ H_4^{\frac5 8}H_0^{\frac 1 8}\left(\frac{dU^2}{f(U)}+U^2
d\Omega_4^2\right)\,,
\\
e^{-\Phi}&=&g_s\left(\frac {H_4}{H_0^3}\right)^{-\frac 1 4}\,,
\quad
f_2=\frac{(2\pi \ell_s)^7 g_s N_0}{\omega_4 V_4} \frac 1{U^4H_0^2}dU\wedge d\tau\,,
\quad
f_4=\frac{(2\pi\ell_s)^3N_c g_s}{\omega_4}\epsilon_4\,,
\end{eqnarray}
where 
\begin{eqnarray}
H_4&=&1+\frac {U_{Q4}^3}{U^3}\,, \quad H_0=1+\frac{U_{Q0}^3}{U^3}\,,
\quad f(U)=1-\frac{U_{KK}^3}{U^3}\,.
\\
\label{eq:UQ0} 
U_{Q0}^3&=&\frac 1 2 \Big(-U_{KK}^3+ \sqrt{U_{KK}^6+\big((
2\pi)^5\ell_s^7g_s \tilde \kappa N_c\big)^2 }\Big)\,,
\\
U_{Q4}^3&=&\frac 1 2 \Big(-U_{KK}^3 +
\sqrt{U_{KK}^6+(2\pi)^2\ell_s^6g_s^2N_c^2}\Big)\,.
\end{eqnarray}
$d\Omega_4$, $\epsilon_4$, and $\omega_4=8\pi^2/3$ are the line
element, the volume form and the volume of a unit $S^4$. $U_{KK}$ is the
coordinate radius of the horizon,
and $V_4$ the volume of D4-brane. $N_0$ and $N_c$ are the numbers
of D0- and D4-branes, respectively.  D0-branes are smeared in the
$x^0,\dots, x^3$ directions.  So $N_0/V_4$ is the number density of
the D0-branes. In order to keep the backreaction of the D0-brane, we require $N_0$ to be
of order $N_c$ as in  \cite{Liu:1999fc} and define $\tilde
\kappa=N_0/(N_c V_4)$ which is $O(1)$ in the large $N_c$.

By changing to the string frame, making the double wick rotation
and taking the field limit $\alpha'\to0$ with $U/\alpha'$ and
$U_{KK}/\alpha'$ finite, the metric
becomes
\begin{eqnarray}
ds^2&=&\left(\frac U R\right )^{3/2} \left(H_0^{1/2}(U) \eta_{\mu\nu}dx^\mu
dx^\nu+ H_0^{-1/2}(U) f(U)d\tau^2\right )
\nonumber \\&&+H_0^{1/2}\left(\frac R
U\right)^{3/2}\left(\frac 1{f(U)} dU^2+U^2 d\Omega_4^2\right)\,,
\end{eqnarray}
and the dilaton
\begin {eqnarray}
e^{\Phi}&=&g_s\left(\frac U R\right)^{3/4}H_0^{3/4}\,,
\end{eqnarray}
where $dx^2= -(dx^0)^2+(dx^1)^2+\cdots+(dx^3)^2$ and $R^3\equiv \pi
\alpha'^{3/2} g_sN_c $ is the limit of $U_{Q4}^3$.
In fact, the
metric is a bubble geometry and the space-time 
ends at $U=U_{KK}$.
To avoid the conical singularity, the period of $\tau$
should be 
\begin{eqnarray}
\beta &=& \frac {4\pi} 3 U_{KK}^{-1/2}R^{3/2} b^{1/2}\,, \quad b\equiv
H_0(U_{KK})
\label{eq:beta}
\end{eqnarray}

 In the open string description, the low energy excitations on D4-brane
are described by a five-dimensional $U(N_c)$ gauge field theory
in the world volume of the D4. By relating the D4-brane tension and the five-dimensional Yang-Mills
coupling constant $g_5$ and compactifying the five-dimensional theory to
four dimensions on the $\tau$ direction, we obtain the four-dimensional Yang-Mills coupling,
\begin{eqnarray}
{g_{YM}^2}=\frac{g_5^2}{\beta}=\frac{4\pi^2 g_s
\ell_s} \beta\,,
\end{eqnarray}
and $b$ and $R^3$ can then be expressed as 
\begin{eqnarray}
b&=& \frac 1 2 (1+ (1+ C \beta^2)^{1/2})\,,\quad
C\equiv(2\pi\ell_s^2)^6 \lambda^2\tilde \kappa^2/U_{KK}^6\,,
\label{eq:H0}
\\R^3&=& \frac {\beta\lambda 
\ell_s^2}{4\pi}\,,
\label{eq:R3}\end{eqnarray}
where $\lambda=g_{YM}^2 N_c$ is the 't Hooft coupling. We can then
define a Kaluza-Klein (KK) mass scale $M_{KK}= 2\pi/\beta$, beyond which
the KK modes will come into play in the four-dimensional gauge theory.
Since we have  imposed the antiperiodic condition on fermions, the
fermions and scalars are also massive with masses at the KK mass scale.  Below
$M_{KK}$, the four-dimensional low-energy theory is a  pure Yang-Mills
theory.

\begin{figure}[!t]
\centerline{\includegraphics[width=7cm]{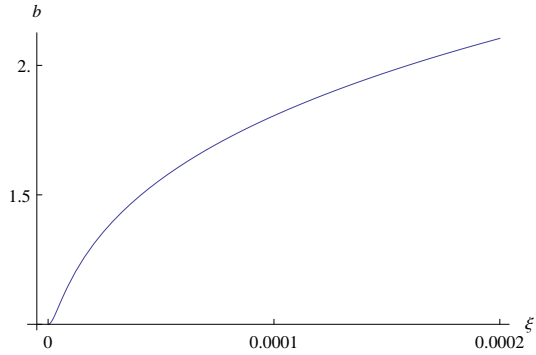}}
\caption{The relation between $b$ and parameter $\xi$.}
\label{fg:H0-xi}
\end{figure}

From (\ref{eq:R3}) and (\ref{eq:beta}) we have 
\begin{equation}
\beta =\frac {4\pi\lambda \ell_s^2} {9  U_{KK}} b\,,\quad
M_{KK}=\frac 9 2 \frac {U_{KK}}{\lambda \ell_s^2 b}\,.
\label{eq:beta-H0}
\end{equation}
Since $b\ge 1$, $U_{KK}\ge 2\lambda\ell_s^2M_{KK}/9$.
From (\ref{eq:beta-H0}) and (\ref{eq:H0}), $\beta$ can be solved 
\begin{eqnarray}
\beta&=&\frac { 4\pi \lambda \ell_s^2}{9 U_{KK}} \frac 1{1-\frac{(2\pi
\ell_s^2)^8}{81 U_{KK}^8}\lambda^4\tilde \kappa^2}
\,,
\end{eqnarray}
and comparing with
(\ref{eq:beta-H0})
we have 
\begin{eqnarray}
b=\frac 1{1-\frac{(2\pi
\ell_s^2)^8}{81 U_{KK}^8}\lambda^4\tilde \kappa^2}\,.
\label{eq:beta-M}
\end{eqnarray}
Since $b>0$, this gives a constraint for $\tilde\kappa$, 
\begin{equation}
|\tilde \kappa|
\leq \frac{9U_{KK}^4}{ (2\pi \ell^2_s)^4\lambda^2}=\frac{\lambda^2 M_{KK}^4
b^4}{9^3\pi^4}.\label{eq:bound}
\end{equation}
If we fix $\beta$, $\lambda$, from (\ref{eq:beta-H0}), $U_{KK}$
goes the same as $b$. And together with (\ref{eq:beta-M}),
$b$ and $\tilde \kappa$ can be
related 
\begin{eqnarray}
b^8-b^7=\frac{9^6\pi^8\tilde \kappa^2}{\lambda^4
M^8_{KK}}=9^6\pi^8\xi^2\,.
\label{eq:H0-kappa}
\end{eqnarray}
For future convenience, we have defined a dimensionless quantity
$\xi$:
\begin{eqnarray}
\xi \equiv \frac{|\tilde{\kappa}|}{\lambda^2 M_{KK}^4}\,.
\label{eq:xi}
\end{eqnarray}
Since we fix $\lambda$ and $M_{KK}$, variation of $\tilde \kappa$ is
equivalent to alteration of $\xi$. 
The left-hand side of (\ref{eq:H0-kappa}) is a monotonic function  increasing from zero for
$b\geq 1$. So for each $\tilde \kappa$, there
is only one solution of $b$, going up as $\tilde \kappa$
increases (see Figure \ref{fg:H0-xi}), and
$U_{KK}$ is similar. So the dependence on $\tilde \kappa$ can be
represented as the dependence on $b$. Since we are interested in the region with
$\lambda\gg 1$, if we choose $\lambda \sim 100$ (or $10^3$) and
 $|\tilde\kappa|<M_{KK}^4$ in order for massive particles to decouple, $\xi$ should be within
$0<\xi<10^{-4}(10^{-6})$ and  the
corresponding $b$ falls in $1<b<1.81(1.005)$. If we allow for the
massive 
modes to be taken into account, $b$ can be chosen in larger domains:
e.g., for $|\tilde
\kappa|<(2M_{KK})^4$,  the corresponding domain of $b$ is $1<b<3.41$
(for $\lambda=10^2$).

This string theory background actually introduces another free
parameter $\tilde\kappa$ into 
the S-S model and is not dual to the vacuum state of the gauge
theory. The same as in \cite{Liu:1999fc}, the dual state may describe some excited
state with (a stochastic averaging of) a constant homogeneous field strength background which
gives the expectation
value of $tr (F_{\mu\nu}\tilde F^{\mu\nu})$
\begin{equation}
\langle tr (F_{\mu\nu}\tilde F^{\mu\nu})\rangle =8\pi^2N_c \tilde
\kappa\,.\label{eq:condensate}
\end{equation}
Since the four-dimensional space-time translation invariance and proper Lorentz
invariance are preserved  in the string background
solution,  we suppose the dual state is a stochastic average over the
background fields in all directions so that the $\langle F\rangle$ is
still zero. Obviously the $P$ and $CP$ invariances are
violated. This is just  similar to the situation in 
 \cite{Liu:1999fc} by H. Liu {\it et al}. Unlike the D(-1)-D3
background used in \cite{Liu:1999fc}, the background here is not
supersymmetric. The self-duality of the field strength may be related
with the supersymmetry. So we cannot say much about the self-duality
of the field strength. 
This state is not the vacuum state of the gauge theory, since in the true vacuum
state, $\theta$ should be zero and there is no $\langle tr (F\tilde F)
\rangle $ condensate (as an abuse of terminology, we use ``condensate"
to denote the expectation value of $tr (F\tilde F) $
not only in vacuum state but also in the excited state).  
We assume that there could exist such excited states in the
corresponding gauge theory and we are interested in the hadron
properties in this kind of states. Such states were proposed to have
some possibilities of being created in the heavy ion collisions
as we stated in the introduction section. In \cite{Wu:2013zxa}, we
have put the Sakai-Sugimoto model in this background and discussed the
effects of $\tilde \kappa$ on the properties of the low energy
Goldstones and mesons. In \cite{Seki:2013nta},
chiral condensate is studied
with finite baryon density in this background. In the next few sections, we will 
 study the $\tilde \kappa$
dependence of the baryon mass spectrum in the S-S model.  

Now we have some independent parameters on the gravity side: $R^3$,
$U_{Q0}^3 $, $ U_{KK}$ and  $g_s$, and $\ell_s$ will be cancelled out
in the final physical results. We also have some parameters on the
gauge theory side $N_c$, $M_{KK}$, $\lambda$ and $\tilde \kappa$.  We
have seen that $\tilde \kappa$ can be related to $b$ and we
can use $b$ to represent $\tilde \kappa$. The final results
on the gauge theory side can be expressed using $N_c$, $M_{KK}$,
$\lambda$ and $b$. We collect the relations here:
\begin{eqnarray}
R^3=\frac {\lambda \ell_s^2}{2M_{KK}}\,,\quad g_s=\frac
{\lambda}{2\pi M_{KK}N_c\ell_s}\,,\quad U_{KK}=\frac 2 9 M_{KK}\lambda
\ell_s^2 b\,. 
\label{eq:constants}
\end{eqnarray}
We fix the gauge theory parameters $M_{KK}$, $N_c$ and $\lambda$, and then
vary $\tilde \kappa$ to see the effects in the hadron physics. This corresponds to fixing the parameters on
the
gravity side: $R^3$, $g_s$, $b/U_{KK}$, and altering
$b$ or $U_{KK}$.

Similar to the discussion of the D4-soliton background
\cite{Kruczenski:2003uq} in the S-S model, in \cite{Wu:2013zxa} we
also discussed the reliability of the background. We simply state the
result here: the valid region of the string theory solution really
corresponds 
to the strong coupling region of the four-dimensional gauge theory 
in the 't Hooft limit.

\section{Classical soliton solution\label{sect:classical}}

After adding $N_f$ D8-anti-D8 branes into the D0-D4 system, we
have introduced $N_f$ flavors and the chiral symmetry $U_L(N_f)\times
U_R(N_f)$. The embedding is nontrivial in $\tau-U$ direction,
$U(\tau)$. The separated D8 and anti-D8 far away combine
near the horizon which corresponds to the  spontaneously breaking of $U_L(N_f)\times
U_R(N_f)$ symmetry to $U_V(N_f)$ in the field theory and  this is
verified by the appearance of massless Goldstones\cite{Wu:2013zxa}. 
We still choose the antipodal embedding for simplicity of the discussion.
Table \ref{tab:config} illustrates the brane configurations.
\begin{table}
\begin{tabular}{c|cccccccccc}
\hline
&0&1&2&3&$4(\tau)$&$5(U)$&6&7&8&9\\
\hline
D4&-&-&-&-&-& &&&&
\\
D8&-&-&-&-&&-&-&-&-&-\\
D0&$=$&$=$&$=$&$=$&-&&&&&\\
\hline
\end{tabular}
\caption{The brane configurations: ``$=$'' denotes the smeared
directions, ``-'' denotes the world volume directions.\label{tab:config}}
\end{table}
The induced metric is 
\begin{eqnarray}
ds^2&=&\left(\frac U R\right )^{3/2}  H_0(U)^{-1/2}\left( f(U)
+\left(\frac R
U\right)^{3}\frac {H_0(U)}{f(U)} {U'}^2\right )d\tau^2
\nonumber \\
&&+\left(\frac U R\right )^{3/2} H_0^{1/2}(U) \eta_{\mu\nu}dx^\mu
dx^\nu+H_0^{1/2}(U)\left(\frac R
U\right)^{3/2}U^2 d\Omega_4^2\,.
\end{eqnarray}
$U'$ is the derivative with respect to $\tau$. We then make the change
of the coordinate $U^3=U_{KK}^3+U_{KK}r^2$,
\begin{eqnarray}
y=r\cos\theta\,, \quad z=r\sin\theta.
\end{eqnarray}
Then D8 is embedded at $y=0$ and on D8 $U^3=U_{KK}^3+ U_{KK} z^2$.
Now the induced metric  becomes
\begin{align}
ds_{D8}^2=&H_0^{1/2}(U)\left(\frac R
U\right)^{-3/2}dx^2+H_0^{1/2}(U)\left(\frac R
U\right)^{3/2}U^2d\Omega_4^2 \nonumber
\\
 &+\frac 4 9 \left(\frac R U\right)^{3/2}\frac
{H_0(U_{KK})}{H_0^{1/2}(U)}(1-h(r)z^2)dz^2
\end{align}

To discuss the baryons spectrum in the S-S model in the D0-D4
background, we adopt the approach of \cite{Hata:2007mb}. The baryon excitations can be viewed as instanton
configurations of the gauge field excitations in the $1,2,3,4,z$
directions  on the D8-branes. By quantizing the collective
coordinates of the instanton, one can obtain the baryon spectrum in
presence of $\tilde \kappa$.
 
We start with the low energy effective action of the non-Abelian gauge
field on the D8-brane. This action
contains two parts: one is from the DBI action
\begin{eqnarray}
S_{\rm YM} = -\tilde{T}U^{-1}_{KK} \int d^4xdz\
2\,H_0^{1/2}(U) {\rm tr}\left[
\frac{1}{4} \frac{R^3}{U} {\cal F}_{\mu\nu}{\cal F}^{\mu\nu} +
\frac{9}{8} \frac{U^3}{U_{KK}} {\cal F}_{z\mu}{\cal F}^{z\mu} \right]
\label {eq:nonabel-DBI0}
\end{eqnarray}
where
\begin{eqnarray}
{\tilde{T}} =
\frac{(2\pi\alpha')^2}{3g_s}\, T_8\, \omega_4\, U^{3/2}_{KK}\, R^{3/2}
= \frac{M^2_{KK} \lambda N_c {b}^{3/2}}{486\pi^3 }
\end{eqnarray}
and the other is from the Chern-Simons terms
\begin{eqnarray}
S_{CS}&=&\frac {N_c}{24\pi^2}\int _{M^4\times R}\omega_5({\mathcal A})\,,
\\
\omega_5({\cal A})&=&{\rm tr}\left({\cal A}\wedge {\cal F}\wedge {\cal
F}-\frac 1 2 {\cal A}^3\wedge {\cal F}+\frac 1 {10} {\cal
A}^5\right)\,.
\end{eqnarray}
The gauge field $\cal A$ and field strength $\cal F$  can be
decomposed to a  $SU(N_f)$
part and a $U(1)$ part:
\begin{eqnarray}
\mathcal{A}&=&\mathcal{A}_\mu
dx^\mu+\mathcal{A}_zdx^z=A+\frac{1}{\sqrt{2N_f}}\hat{A}=A^aT^a+\frac{1}{\sqrt{2N_f}}\hat{A}\,,\nonumber
\\
\mathcal{F}&=&d\mathcal{A}+i\mathcal{A}\wedge\mathcal{A}=F+\frac{1}{\sqrt{2N_f}}\hat{F}\,.
\label{eq:F}
\end{eqnarray}
To simplify the calculation, we make the replacement $z\to z U_{KK}$,
$ \mathcal{A}_z\to \mathcal{A}_z/U_{KK}$, $x_\mu\to x_\mu/M_{KK}$,
$\mathcal{A}_\mu\to \mathcal{A}_\mu
M_{KK}$ to work with dimensionless $x$, $z$ and $\mathcal{A}_\mu$, $\mathcal{A}_z$. The
Chern-Simons terms are not affected and the Yang-Mills action changes
to
\begin{align}
S_{\rm YM} &= - \tilde{T}M^{-2}_{KK}\frac 9{4b} \int d^4xdz\
H_0^{1/2}(U) {\rm tr}\left[
\frac{1}{2} \frac{U_{KK}}{U} {\cal F}_{\mu\nu}{\cal F}^{\mu\nu} +
 \frac{U^3}{U^3_{KK}}b {\cal F}_{z\mu}{\cal F}^{z\mu} \right], 
\\&=-a\lambda N_c b^{1/2} \int d^4xdz\
H_0^{1/2}(U) {\rm tr}\left[
\frac{1}{2} K(z)^{-1/3} {\cal F}_{\mu\nu}{\cal F}^{\mu\nu} +
 K(z)b {\cal F}_{z\mu}{\cal F}^{z\mu} \right],\label {eq:nonabel-DBI1}
\end{align}
where $a=\frac 1{216\pi^3}$, and  $K(z)=1+z^2$. To use $U_{KK}$ to make $z$ and $A_{z}$
dimensionless is to keep the coefficient of the second term 
finite in the $\ell_s\to0$ limit. We can then make another change of
coordinate $z\to b^{1/2}z$, $A_z\to A_z/b^{1/2}$, to give the standard
normalization of the second term:
\begin{align}
S_{\rm YM} &=-a\lambda N_c b \int d^4xdz\
\,H_0^{1/2}(U) {\rm tr}\left[
\frac{1}{2} K(zb^{1/2})^{-1/3} {\cal F}_{\mu\nu}{\cal F}^{\mu\nu} +
 K(zb^{1/2}) {\cal F}_{z\mu}{\cal F}^{z\mu} \right],\label {eq:nonabel-DBI2}
\end{align}
The integrand is different from the original one in \cite{Hata:2007mb} by $b$ factors
inside $K$ and the overall factor $H^{1/2}_0(U)$.

By the same reasoning  as in \cite{Hata:2007mb}, since we are working in the large $\lambda$
region, we can make a $1/\lambda$ expansion. It is convenient to make
another rescaling of the coordinates and ${\cal A}_M$,
\begin{eqnarray}
x^0\to x^0&,&\quad x^M\to\lambda^{-1/2} x^M\nonumber\\
\mathcal{A}_0(t,x)\to{\mathcal A}_0(t,x)&,&\quad
\mathcal{A}_M(t,x)\to\lambda^{1/2}{\mathcal A}_M(t,
x)\nonumber\\
\mathcal{F}_{0M}(t,x)\to\lambda^{1/2}{\mathcal F}_{0M}(t,x)&,&\quad
\mathcal{F}_{MN}(t,x)\to\lambda{\mathcal
F}_{MN}(t,x)\,,
\end{eqnarray}
where $M,N=1,2,3,z$. Expanding the Lagrangian with respect to $1/\lambda$,
and  keeping terms
to $O(\lambda^0)$, we have
\begin{align}
S_{\rm YM}=-aN_cb^{3/2}\int d^4xdz
&\left[\frac{\lambda}{4}(F^a_{MN})^2-\frac{b\,z^2}2(\frac{5}{12}-\frac{1}{4b})(F^a_{ij})^2+\frac{b\,z^2}{4}(1+\frac{1}{b})(F^a_{iz})^2-\frac
1 2(F^a_{0N})^2\right.\nonumber\\
&\left.+\frac{\lambda}{4}\hat
F^2_{MN}-\frac{b\,z^2}2(\frac{5}{12}-\frac{1}{4b})\hat
F^2_{ij}+\frac{b\,z^2}{4}(1+\frac{1}{b})\hat F^2_{iz}-\frac 1 2 \hat
F^2_{0N}+O(\lambda^{-1})\right]
\label{eq:S-YM}
\end{align}
with $i=1,2,3$.  For simplicity we are now working with only two
flavors, that is $N_f=2$. We
have  decomposed the gauge field into the $SU(2)$ and $U(1)$ parts as in
(\ref{eq:F}). The Chern-Simons action can also be decomposed as
\begin{align}
S_{CS}=&\frac{N_c}{24\pi^2}\epsilon_{MNPQ}\int d^4xdz[\frac{3}{8}\hat{A_0}tr(F_{MN}F_{PQ})-\frac{3}{2}\hat{A}_M tr(\partial_0A_NF_{PQ})\nonumber\\
&+\frac{3}{4}\hat{F}_{MN}tr(A_0F_{PQ}) +\frac{1}{16}\hat{A}_0\hat{F}_{MN}\hat{F}_{PQ}-\frac{1}{4}\hat{A}_M\hat{F}_{0N}\hat{F}_{PQ}+(...)]
\label{eq:S-CS}
\end{align}
The ellipsis denotes some total derivative terms.
The EOM for the gauge fields can then be obtained up to
$O(\lambda^{-1})$:
\begin{eqnarray}
D_MF_{MN}=0\,, 
\quad D_MF^a_{0M}+\frac{\epsilon_{MNPQ}}{64\pi^2ab^{3/2}}\hat
F_{MN}F^a_{PQ}=0\label{eq:EOM-SU2}\\
\partial_M \hat F_{MN}=0\,, \quad \partial_M\hat
F_{0M}+\frac{\epsilon_{MNPQ}}{128\pi^2ab^{3/2}}
F^a_{MN}F^a_{PQ}=0
\label{eq:EOM-U1}
\end{eqnarray}
$\frac 1 2 \hat A_0$ is coupled to the quark number
operator and
the instanton number $n=\frac 1 {32\pi^2}\epsilon_{MNPQ}\int
d^4xtr(F_{MN}F_{PQ})$ is just the baryon number. The BPST 
one-instanton solution of the EOM of $F_{MN}$ is equivalent to the
Skyrmion
description of a baryon,
\begin{align}
&A_M=-if(\xi)g^{-1}(x)\partial_M g(x),\nonumber  \\
f(\xi)=\frac{\xi^2}{\xi^2+\rho^2},\quad &\xi^2=(\vec x-\vec
X)^2+(z-Z)^2,\quad 
g(x)=\frac 1 \xi((z-Z)  -i(\vec x-\vec X)\cdot \vec \sigma)\,,
\end{align}
where $\vec\sigma$ are the Pauli matrices. The EOM of the $U(1)$ $\hat A_M$ gives the
solution $\hat A_M=0$ up to a gauge
transformation. The equation for $A_0$ then becomes $D_M^2A_0=0$. The
same as the argument in \cite{Hata:2007mb}, the solution with
vanishing boundary condition at infinity is given by $A_0=0$.
Inserting the BPST solution into the EOM of $\hat A_0$, we have 
\begin{align}
-\partial_M^2\hat A_0+\frac {3\rho^4}{\pi^2ab^{3/2}(\xi^2+\rho^2)^4}=0
\end{align}
and the solution is 
\begin{align}
\hat A_0=-\frac 1 {8\pi^2ab^{3/2}}\frac
{\xi^2+2\rho^2}{(\rho^2+\xi^2)^2}\,.
\label{eq:A0}
\end{align}
Substituting this solution into the action,  we obtain the soliton mass
from the on-shell
action $S=-\int dt M$:
\begin{eqnarray}
  M&=& 8\pi^2ab^{3/2}
N_c\left(\lambda+\frac{1}{12}
(3-b)(2Z^2+\rho^2)+\frac{1}{320\pi^4\rho^2a^2b^{3}}+O\Big(\frac 1
\lambda\Big)\right)
\label{eq:M}
\end{eqnarray}
If $b<3$, we can find out the minimum of $M$ at $\rho_{min}$ which
characterize the size of the baryon:
\begin{eqnarray}
\rho^2_{min}=\frac 1 {4\pi^2}\sqrt{\frac 3 5} \frac 1 {a b^{3/2} \sqrt
{3-b}}\,.\label{eq:rhomin}
\end{eqnarray}
\begin{figure}
\includegraphics[width=10cm]{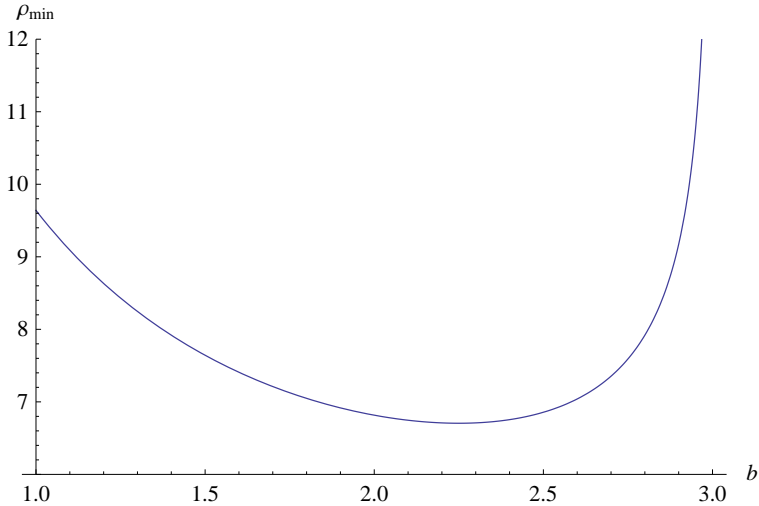}
\caption{The change of the size of the baryon $\rho_{min}$ with
respect to $b$. \label{fig:rhomin}}
\end{figure}
Inserting $\rho_{min}$ into $M$, we have the minimum of the
soliton mass,
\begin{align}
M_{min}=8\pi^2ab^{3/2}
N_c\left(\lambda+9\pi\sqrt{\frac 3 5} \frac{\sqrt{
3-b}}{b^{3/2}}+ \frac 1 6(3-b)Z^2+O\Big(\frac 1
\lambda\Big)\right)\,.\label{eq:M-min}
\end{align}
 From figure (\ref{fig:rhomin}) we see that
the size goes down first as $b$ increases and grows and blow up as
$b\to 3$. We define $b_{min}=9/4$ at which $\rho_{min}$ is the minimum. 
At $b=3$,  the second term of (\ref{eq:M-min}) vanishes and it is obvious that
$\rho_{min}$ goes to infinity if we ignore $O(1/\lambda)$ terms.

\begin{figure}
\includegraphics[width=6cm]{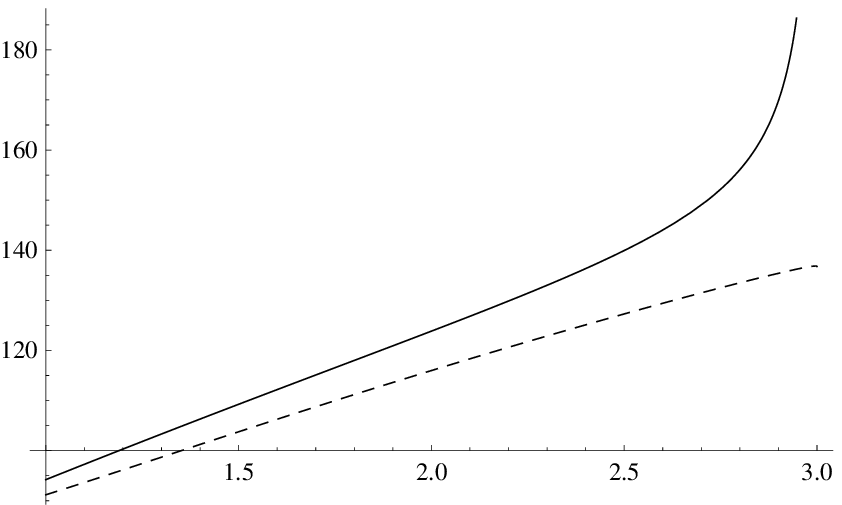}
\includegraphics[width=6cm]{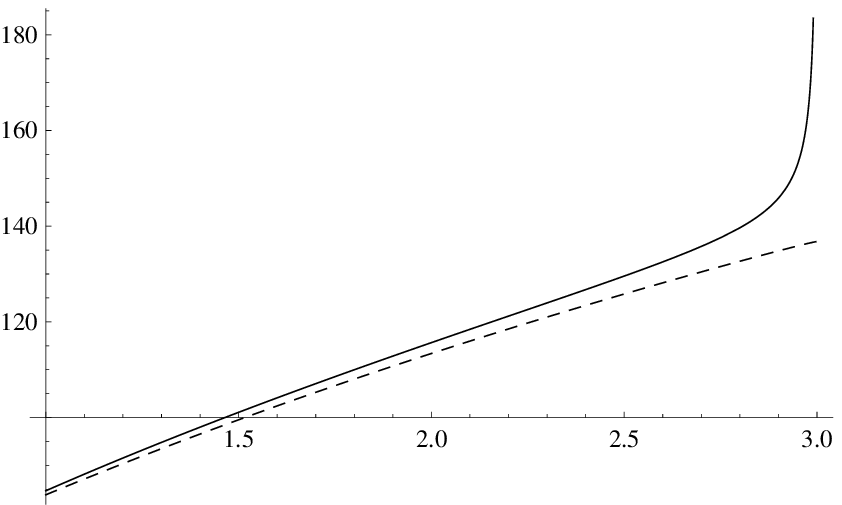}
\caption{The integral of $S_{\rm YM}$: the solid line denotes the one
not expanded with respected to $1/\lambda$
and the dashed line the one  expanded to $O(\lambda^0)$. The left figure is for
$\lambda=100$ and the right one for $\lambda=250$. \label {fig:int} }
\end{figure}

However, we have to estimate the region of $b$ for which the
$1/\lambda$ expansion of the action to $O(\lambda^0)$ is good.
The $O(\lambda^0)$ terms, i.e., the second and the third terms in the
bracket in  (\ref{eq:M-min}),
are  monotonically decreasing for $1<b<3$.
So  if for $b=1$ the $O(\lambda^0)$ term is smaller than
$\lambda$, it will be so for all $1<b<3$. We can estimate the
value:
at $b=1$, for $Z=0$, the second term is $9\sqrt{\frac 65}\pi\simeq
31$. So it is enough to choose the $\lambda$ to be larger than $10^2$
to have  $O(\lambda^0)$ terms smaller than the $O(\lambda)$ term. From (\ref{eq:xi}) and (\ref{eq:H0-kappa}), we have a
constraint for $\xi$ and $b$, $\xi<10^{-4}$ for $\lambda=100$ and
$1<b<1.81$ for $\tilde \kappa<M_{KK}^4$ if we do
not consider massive modes. However, 
for fixed $\lambda$, as $b$ approaching 3, $\rho_{min}^2$ goes to infinity and larger than
$\lambda$, and the $1/\lambda$ expansion may not be a good approximation at such a large 
$\rho$. It is hard to estimate the range $b$ for which the expansion
is good.
We will use an operative estimation as follows. The second term which
is divergent for  $\rho\to \infty$ in the bracket of (\ref{eq:M}) comes
from the integral of the  static BPST solution. So  we could
evaluate $S_{\rm YM}$ in (\ref{eq:nonabel-DBI2}) using BPST
solution  numerically at $\rho_{min}$ with $\hat A_0=0$ and
compare it with (\ref{eq:S-YM}) which is the $1/\lambda$
expansion of the same integral. See figure \ref{fig:int} for
illustration of the differences. We can see that near $b=3$ the
differences become divergent because $\rho_{min}\to
\infty$ for $b\to 3$.  Since the evaluation of the integration
(\ref{eq:nonabel-DBI2}) partly takes account of higher order 
$O(1/\lambda)$
contributions, we would expect the difference should be smaller than the
$O(\lambda^0)$ terms in order to have a good approximation. This
requirement puts a constraint on the range of $b$. For $\lambda\sim 10^2$, this will give a rough
constraint  $1<b<1.5$ which is within the above  $1<b<1.81$.
For $\lambda\sim 250$, this will give a constraint about $1<b<1.8$.
But  (\ref{eq:xi}) and (\ref{eq:H0-kappa}) give a constraint
$1<b<1.25$. In
$1.25<b<1.8$, we would expect that the massive modes of the gauge
theory  come into play.
In general, equations (\ref{eq:xi}) and (\ref{eq:H0-kappa}) require  that the larger
$\lambda$ is,  the smaller the region for $b$ is for $\tilde
\kappa<M_{KK}^4$. The above
criterion for a good $O(\lambda^0)$ approximation requires the opposite, that is,
the larger $\lambda$ is, the larger the region for $b$ is. 

Above we consider the expansion of $S_{\rm YM}$ from BPST part to be a good
expansion. If we also include the $\hat A_0$ contributions, there are
some subtleties to be considered. Recall that, in \cite{Wu:2013zxa}
when we consider the condition that $e^{\Phi}\ll1$ which is needed for
the suppression of the string loop effects, we need $UH_0(U)\ll
g_s^{-4/3}R$, and from (\ref{eq:constants}) we have
\begin{eqnarray}
\frac U{U_{KK}}=(1+z^2b/\lambda)^{1/3}\ll
\frac{9\pi^{4/3}N_c^{4/3}}{\lambda^2 b}\,.
\label{eq:z-constraint}
\end{eqnarray} 
Though we are working in the
large $N_c$ limit and the right hand side of (\ref{eq:z-constraint})
can be arbitrarily large, to do numerical computation, we have to fix a
finite $N_c$. Then (\ref{eq:z-constraint}) gives a constraint for the
value of $z$ and thus
introduces a cut-off of the integration over $z$. This does not affect the
integral of BPST part since the integral converges quickly such that
the variation of  limits of $z$ integration does not change the
result too much, as long as the integration limits are large enough. But since $\hat F_{0z}^2\sim 1/\xi^6$ for large $z$
and the $O(\lambda^{-1})$ term of $k(z)$ is proportional to  $z^2$, the second term in (\ref{eq:nonabel-DBI1}) does
not have a good convergent property. We have to choose the integration
cut-off  
of $z$ to have a good $1/\lambda$ expansion to $O(\lambda^0)$. So the
integration limits cannot be chosen arbitrarily large. By
numerical test, we find out that the choice for the  limits of $z$ integration to be of
$O(\lambda/10)$ is a good one. And this does not change the
results in previous paragraph. We can then look at how large the $N_c$
should be from (\ref{eq:z-constraint}), that is $N_c\gg\frac
{\lambda^{3/2}b^{3/4}}{3^{3/2}\pi}(1+\frac {z^2b}{\lambda})^{1/4}$. If
we choose $\lambda=10^2$ and
$z=\lambda/10=10$, $N_c\gg 10^2$ for $b\in (1,1.5)$. For
$\lambda\gg10^2$ and $z\sim \lambda/10$, $N_c\gg \frac
{\lambda^{7/4}b}{\sqrt{270}\pi}\sim 0.02 b \lambda^{7/4}$. So $N_c$
should be rather large for the $1/\lambda$ expansion to be good. 

Now we can see that adding the condensate will cause the baryon to
shrink for small $\tilde \kappa$. From (\ref{eq:M}) the
second term is an attractive potential and the third term is a repulsive
one. For small $b$ near $b=1$, the attractive potential [including the
$b^{3/2}$ factor before the bracket in (\ref{eq:M}) ] grows 
while the repulsive one decreases as $b$ increases. So it can be
understood that the baryon size
shrinks as $b$ increases. For larger $b$, the attractive potential 
also decreases and the repulsive one decreases more slowly with
increasing $b$.  At 
$b>b_{min}$ which corresponds to larger $\tilde
\kappa$ , we can always choose $\lambda$ large enough to make the $1/\lambda$ expansion 
 a good approximation to $O(\lambda^0)$. At such $b$, the repulsive force will be
decreasing more slowly than the attractive force and the
radius will be increasing with $b$.
However, in this case,  $\tilde
\kappa$ is always greater than $M^4_{KK}$ for our chosen
$\lambda>100$, and the massive
modes, including fermions, bosons and KK modes from the five-dimensional 
gauge theory, may come into play. 

\section{Quantization of the collective modes and the spectrum for
baryons\label{sect:quantum}}
The BPST instanton solution are parametrized by position $\vec X$,
$Z$,
size $\rho$, and together with the global $SU(2)$ orientation
parametrized by three independent $SU(2)$ parameters, $a_I$, $(I=1,2,3,4)$ with
$\sum_I a_I^2=1$, they form the moduli space of the
one-instanton  solution, $\mathcal{M}=\mathbb{R}^4\times
\mathbb{R}^4/\mathbb{Z}_2$.   As in \cite{Hata:2007mb}, we also define
$y_I$,  $I=1,2,3,4$ to combine 
$\rho$, $a_I$, with $\sum_I y_I^2 =\rho^2$ and
$a_I=y_I/\rho$.  The basic idea to study the baryon
spectrum in the approach of \cite{Hata:2007mb} is to regard the
soliton as a slowly moving particle described by time-dependent
collective coordinates. The motion of the particle can be
quantized, and the energy spectrum can be obtained by the eigenvalues
of the Hamiltonian for the collective coordinates.

First, we promote the coordinates to be time dependent: the positions
$\vec X(t)$, $Z(t)$, the size $\rho(t)$, $SU(2)$ orientation $a_I(t)$, or
collectively denoted by $X^\alpha=(\vec X(t), Z(t),y_I(t))$. The $SU(2)$ gauge
field becomes time dependent,
\begin{align}
A_M(t,x)=VA_M^{cl}(x,X^{\alpha}(t))V^{-1}-iV\partial_MV^{-1}.
\end{align}
where $A_M^{cl}(x,X^\alpha)$ is the BPST instanton solution with $\vec
X$, $Z$, $\rho$ replaced by time-dependent ones. $V(x,t)$ is a $SU(2)$
$2\times 2$ matrix which is asymptotic to $\mathbf{a}=a_4(t)+ia_a(t)
\sigma^a$ at $z\to \infty$, where $\sigma^a$, $a=1,2,3$ are Pauli matrices.
The field strength then becomes
\begin{eqnarray}
F_{MN}=VF_{MN}^{cl}V^{-1}\,,\quad F_{0M}=V(\dot X^\alpha\partial_\alpha
A_M^{cl}-D_M^{cl}\phi)V^{-1}
\end{eqnarray}
where $\Phi=-iV^{-1}\dot V$. The dot denotes the derivative with respect to $t$,
i.e., $\partial_0$, and $\partial_\alpha$ is the derivative with respect to
collective coordinates $X^\alpha$. $D_M$ is the covariant derivative
$D_M=\partial_M+i[A_M^{cl},\ ]$, using BPST solution $A^{cl}_M$.
$F_{MN}$ automatically satisfies the EOM of the first equation in
(\ref{eq:EOM-SU2}). $F_{0M}$ and $\hat F_{0M}$ have to satisfy the
second equations in
(\ref{eq:EOM-SU2}) and (\ref{eq:EOM-U1}). By the analysis in \cite{Hata:2007mb}, the
solution of $A_{0}$ is the same as  in \cite{Hata:2007mb} and we would not
repeat it here. The solution of $\hat A_{0}$ is just (\ref{eq:A0}) with
collective coordinates replaced by the time-dependent ones. 

The motion of the collective coordinates can be
characterized by the Lagrangian up to $O(1/\lambda)$ obtained by substituting the time-dependent solution
into (\ref{eq:S-YM}) and (\ref{eq:S-CS}): 
\begin{eqnarray}
L&=&\frac 1 2 m_X g_{\alpha\beta}\dot X^\alpha \dot X^\beta -U(X^\alpha)
\\
&=& \frac 1 2 m_X \dot {\vec X}^2 +\frac 1 2 m_Z \dot Z^2 +\frac
1 2 m_y \dot y_I\dot y_I -U(X^\alpha)
\end{eqnarray}
where $g_{\alpha\beta }$ is the metric for the instanton moduli $
ds^2=g_{\alpha\beta}dX^{\alpha}dX^{\beta}=d\vec X^2+dZ^2+2dy_I^2$, and  $m_X=m_Z=\frac 1 2
m_y=8\pi^2ab^{3/2}N_c$. The potential $U(X)$ is
the same as (\ref{eq:M}) except that the collective coordinates are
time dependent. The Hamiltonian can then be  obtained:
\begin{eqnarray}
H&=&H_X+H_Z+H_y\,,\nonumber \\
H_X&=&\frac 1 2 m_X \dot X^i\dot X^i +8\pi^2\lambda ab^{3/2}N_c=\frac
1 {2m_X}P_X^2+M_0\,,
\\
H_Z&=&\frac 1 2 m_Z \dot Z^2+\frac {4\pi^2}3
ab^{3/2}(3-b)N_cZ^2=\frac 1 {2m_Z}P_Z^2+   \frac 1 2 m_Z \omega_Z^2  Z^2
\\
H_y&=&\frac 1 2 m_y\dot y_I\dot y_I+\frac {2\pi^2}3
ab^{3/2}(3-b)N_c \rho^2+\frac {N_c}{40\pi^2ab^{3/2}}\frac 1
{\rho^2}
\nonumber \\ 
&=&\frac 1 {2 m_y}P_y^2+\frac 1 2
m_y\omega_y^2\rho^2+\frac Q{\rho^2}\,.
\end{eqnarray}
We have defined $M_0=8\pi^2\lambda ab^{3/2}N_c$, $\omega_Z=\frac 1
3(3-b)$, $\omega_y=\frac 1  {12}(3-b)$ and $Q=\frac
{N_c}{40\pi^2ab^{3/2}}$, and the momentum $P_X$, $P_Z$, $P_y$ follow the
standard definition as the conjugate of the collective coordinates.

We quantize the soliton at rest to obtain the baryon spectrum. The quantization procedure is to replace the momenta in the
Hamiltonian
to the corresponding differential operators which act on the normalizable wave
function describing the motion of the
soliton
\begin{eqnarray}
H_Z&=&-\frac 1 {2m_Z}\frac {\partial^2}{\partial Z^2}+   \frac 1 2 m_Z \omega_Z^2  Z^2
\\
H_y&=&-\frac 1 {2 m_y}\frac{\partial^2}{\partial y_I^2}+\frac 1 2
m_y\omega_y^2\rho^2+\frac Q{\rho^2}\,.
\end{eqnarray}
For the wave function to describe fermions, we need to impose 
antiperiodic boundary condition: $\psi(-a_I)=-\psi(a_I)$.
The Hamiltonian is almost the same as the one in \cite{Hata:2007mb}
only
with redefined $m_Z$, $m_y$, $\omega_Z$, $\omega_y$ and $Q$. So we can
use their result directly,
\begin{eqnarray}
M&=&M_0+E_y+E_Z\,, \quad M_0=\frac {\lambda N_c b^{3/2}}{27\pi}
\\
E_y&=&\omega_y(\tilde
l+2n_\rho+2)=\omega_y(\sqrt{(l+1)^2+2m_yQ}+2n_\rho+1)\nonumber
\\&=&\frac{1}{2\sqrt 3}\sqrt{3-b}\left(\sqrt{(l+1)^2+\frac 4 5
N_c^2}+2n_\rho+1\right)\,,\label{eq:Ey}
\\
E_Z&=&\omega_Z(n_z+\frac 1 2)=\frac{1}{\sqrt 3}\sqrt{3-b}(n_z+\frac 1
2)\label{eq:EZ}
\end{eqnarray}
where $l=1,3,5,\dots$, $n_\rho=0,1,2,3,\dots$, and $n_z=0,1,2,\dots$.
That $l$ takes odd integers is the requirement of the antiperiodic
boundary condition of the wave function.
 
Now we have the baryon mass spectrum. However, there is some
uncertainty in the formula. As noted in \cite{Hata:2007mb} the
$O(N_c^0)$ contribution from $M_0$ is ignored which is the same
$N_c$ order of the  zero-point energy in (\ref{eq:Ey}) and
(\ref{eq:EZ}). And the contributions to the zero-point energy from the heavy massive mode
excitations around
the instanton are also ignored. The sum of these zero-point energy may give a divergent contribution
and need be subtracted by the vacuum energy to make it finite. These
contributions in general will also depend on $b$. So the $b$ dependence of $M_0$
together with all the other zero-point energy contributions cannot be estimated
at this stage. If we insisted the large $N_c$ and large $\lambda$
expansions to be  good approximations, the $O(N_c^0)$ and $O(\lambda^0)$
contribution from the zero-point energy should not be dominant over
$M_0$ which is $O(\lambda N_c)$. However, this may not be realistic
for $N_c=3$ and the real $\lambda$.  
Considering such uncertainty, what we can do is to discuss the mass difference between
baryon excitation states. 

From (\ref{eq:Ey}) and (\ref{eq:EZ}),  we can easily separate
the $b$ dependence in $E_y$ and $E_Z$:
\begin{eqnarray}
M=M_0+\sqrt{\frac{3-b}{2}}\big(E_y(b=1)+E_Z(b=1)\big),\quad
\Delta M=\sqrt{\frac{3-b}{2}}\Delta M(b=1)
\end{eqnarray}
So, the difference between two baryon states
is proportional to $\sqrt{3-b}$. For $1<b<3$, the difference
become smaller as $b$ increases and disappears at $b=3$.  
This result is independent of
the specific properties such as spin or isospin of the baryons.
However, as discussed in section \ref{sect:classical}, the suitable
region of $b$ for our method is constrained by
the $1/\lambda$ approximation. For $\lambda\sim 100$, only $1<b<1.5$
could give a good $1/\lambda$ approximation to $O(1)$ and in this region $\tilde
\kappa<M_{KK}^4$. The massive modes of the gauge theory decouples. For $\lambda$ large enough, $b$ can be
near $3$ with good $1/\lambda$ approximation but $\tilde\kappa$ can be greater than $M^4_{KK}$. So the behavior
of $M$ for such large $b$ has already included the contributions from
the massive modes in the gauge theory. If we are only interested in
the contribution of the massless gauge field, we cannot trust the large $b$
region and cannot say much about the behavior in this region. However, near
$b=1$, the qualitative behavior of result above can be trusted.

\section{Conclusion and discussion\label{sect:discussion}}
We have discussed the baryon spectrum in the S-S model with smeared D0
charge turned on in the D4-soliton background. The D0-D4 background
corresponds to an excited state with $\tilde \kappa$ which is
proportional to a  nonzero $tr(F_{\mu\nu}\tilde F^{\mu\nu})$. The
dependence on $\tilde \kappa$ is through a parameter $b$, which is
monotonically increasing  with $\tilde \kappa$.
We follow the method in \cite{Hata:2007mb}, in which the static baryon is
represented by an instanton solution and by quantizing the collective
coordinates, one can obtain the baryon spectrum. 

In  the classical analysis, for $b<3$, the soliton mass has an attractive
potential and a repulsive potential. For small $\tilde\kappa$, the
attractive potential grows with $\tilde\kappa$ and the repulsive potential decreases. This
causes the size of the baryon to shrink. For larger $\tilde\kappa$, the
massive modes in the gauge theory may also come into play, and the size of
the baryon will grow with $\tilde\kappa$. 
It may be possible that as we increase the $\lambda$, the
upper bound for $b$ goes closer to $3$ according to our criterion, and the radius really grows
larger and larger such that the baryon breaks up. In this case, more and more
KK modes come in and may play a role in driving the baryon
unstable. 
However, our criterion for
the $1/\lambda$ approximation may not be applicable for $b$ very close to
$3$. This is because the integration we used in the criterion only
counts in parts of the $O(1/\lambda)$ contributions, and the other
$O(1/\lambda)$ contributions can be comparably larger than the $O(\lambda^0)$ terms
near $b=3$ in (\ref{eq:M}) in which the $O(\lambda^0)$ terms
is exact zero at $b=3$ for finite $\rho$ and $Z$. This is partly the
reason for the divergence of the radius in (\ref{eq:rhomin}) at $b=3$
to $O(\lambda^0)$. In this case, we cannot say anything near
$b=3$ in our present approximation. Nevertheless, for $b>3$, the
$O(\lambda^0)$ term is negative and nonzero which means a repulsive
potential. We can choose large
enough $\lambda$ so that $O(\lambda^0)$ dominate the higher order
terms at finite $\rho$ and $Z$. Obviously, the baryon cannot be 
formed with only repulsive
force. So, this also indicates a possibililty that for large enough
$\lambda$ and $b$ where the massive KK modes contribute, the baryons cannot
exist.
 
In the analysis of the baryon
spectrum by quantization of collective coordinates, the zero-point energy
 cannot be estimated by the present method. So what we can consider is the
difference between baryon masses. The dependence of the mass
difference on the $\tilde \kappa$ is simply $\sqrt{\frac{3-b}2}\Delta
M(b=1)$  and  the factor is
independent of the spin and isospin. So with $\tilde \kappa$ turned on
the mass difference between baryons will be smaller.
For $b>3$, the baryon masses become complex which also indicates
that they cannot be stable. This is not surprising since  the
repulsive potential in the classical analysis of the previous
paragraph is used here in the Sch\"odinger equation. 
So, as long as the large $\lambda$ expansion is applicable for some
region in $b>3$,
we conclude that the baryons cannot exist in this region. However,
at this region $\tilde \kappa >M_{KK}^4$, contribution from the
massive KK modes of the gauge theory should be counted in.
Though this may not be the property of the low energy QCD, it is a
property of the world volume theory of the D4 compactified on a circle. 

As we learned from the study in \cite{Wu:2013zxa},   the mass spectra of the mesons
and the interactions between mesons do depend on $\tilde\kappa$, which
means that $\tilde\kappa$ really affects the interactions between
quarks.  But this effect is not so drastic as to endanger the
stability of mesons there; however, in the present paper strong
$\tilde\kappa$ does destroy baryons. This may be qualitatively
explained as follows: a baryon is formed by $N_c$ quarks while a meson
is only formed by a quark-antiquark pair. Since we are discussing in
the large $N_c$ region, a baryon may consist of a large number of
quarks.  It is not surprising that the total effect of
$\tilde\kappa$ on the interactions among such a large number of quarks
should be much larger than the effect on only one pair of quark and
antiquark which forms a  meson, and even large enough to destabilize
the baryons. As for the chiral symmetry breaking, it is characterized
by the $\bar q q$ condensate, which involves only two quark fields
$q$. So similar to the meson cases, from the study in \cite
{Wu:2013zxa}, $\tilde \kappa$ effect is too
small to demolish chiral symmetry breaking.  From a
geometric point of view in the S-S model, the spontaneous chiral
symmetry breaking is realized as the merging of the D8-brane and
anti-D8-brane. So, as long as this kind of geometry is present and it
is a valid dual description of the field theory, the
spontaneous chiral symmetry breaking is expected to be there. For the
geometry to be valid, $b$ cannot be arbitrarily large which can be
seen from (\ref{eq:z-constraint}) for fixed $N_c$ and $\lambda$.

As in most gauge-gravity dual analysis, our analysis is done in the
large $N_c$ limit and large 't Hooft coupling region. Since the
baryon spectrum demonstrates the right large $N_c$ behavior as in  \cite{Hata:2007mb}, we
expect that the model also captures the qualitative $\tilde\kappa$ behavior at least for small $\tilde \kappa$ for QCD-like theory at large $N_c$ with large $\lambda$. We have
estimated that $\lambda$ should roughly be larger than $10^2$, and $N_c$ should also be rather
large according to the discussion in section \ref{sect:classical}. This also
applies to the original discussion on baryons in the S-S model  with
$\tilde\kappa=0$ in \cite{Hata:2007mb}.  
Although in \cite{Hata:2007mb}, the
authors did compared the result with the experiments with $N_c=3$
heuristically, it 
should be understood that it is still a long way to the realistic $N_c$
and $\lambda$.

\section*{Acknowledgements}
This work is supported by the NSF of China under Grant No.11105138 and
11235010 
and is also supported by the Fundamental Research Funds for the
Central Universities under Grant No.WK2030040020. We also thank Da
Zhou 
for helpful discussion.


\begin{thebibliography}{99}
%\cite{Kharzeev:1998kz}
\bibitem{Kharzeev:1998kz}
  D.~Kharzeev, R.~D.~Pisarski and M.~H.~G.~Tytgat,
  %``Possibility of spontaneous parity violation in hot QCD,''
  Phys.\ Rev.\ Lett.\  {\bf 81} (1998) 512
  [hep-ph/9804221];
  %%CITATION = HEP-PH/9804221;%%
  %181 citations counted in INSPIRE as of 18 Mar 2013
%\cite{Kharzeev:1998kya}
%\bibitem{Kharzeev:1998kya}
%  D.~Kharzeev, R.~D.~Pisarski and M.~H.~G.~Tytgat,
  %``Parity odd bubbles in hot QCD,''
  hep-ph/9808366;
  %%CITATION = HEP-PH/9808366;%%
  %10 citations counted in INSPIRE as of 18 Mar 2013
%\cite{Kharzeev:2000na}
%\bibitem{Kharzeev:2000na}
%  D.~E.~Kharzeev, R.~D.~Pisarski and M.~H.~G.~Tytgat,
  %``Aspects of parity, CP, and time reversal violation in hot QCD,''
 %Submitted to: Int.J.Mod.Phys.A
  [hep-ph/0012012].
  %%CITATION = HEP-PH/0012012;%%
  %26 citations counted in INSPIRE as of 18 Mar 2013


%\cite{Buckley:1999mv}
\bibitem{Buckley:1999mv}
  K.~Buckley, T.~Fugleberg and A.~Zhitnitsky,
  %``Can theta vacua be created in heavy ion collisions?,''
  Phys.\ Rev.\ Lett.\  {\bf 84} (2000) 4814
  [hep-ph/9910229].
  %%CITATION = HEP-PH/9910229;%%
  %37 citations counted in INSPIRE as of 18 Mar 2013

%\cite{Kharzeev:2004ey}
\bibitem{Kharzeev:2004ey}
  D.~Kharzeev,
  %``Parity violation in hot QCD: Why it can happen, and how to look
  %for it,''
  Phys.\ Lett.\ B {\bf 633} (2006) 260
  [hep-ph/0406125].
  %%CITATION = HEP-PH/0406125;%%
  %193 citations counted in INSPIRE as of 22 Mar 2013

%\cite{Shuryak:2001jh}
\bibitem{Shuryak:2001jh} 
  E.~V.~Shuryak and A.~R.~Zhitnitsky,
  %``Domain wall bubbles in high-energy heavy ion collisions,''
  Phys.\ Rev.\ C {\bf 66}, 034905 (2002)
  [hep-ph/0111352].
  %%CITATION = HEP-PH/0111352;%%
  %10 citations counted in INSPIRE as of 02 Aug 2014
%\cite{Kharzeev:2007jp}
\bibitem{Kharzeev:2007jp} 
  D.~E.~Kharzeev, L.~D.~McLerran and H.~J.~Warringa,
  %``The Effects of topological charge change in heavy ion collisions: 'Event by event P and CP violation',''
  Nucl.\ Phys.\ A {\bf 803}, 227 (2008)
  [arXiv:0711.0950 [hep-ph]].
  %%CITATION = ARXIV:0711.0950;%%
  %514 citations counted in INSPIRE as of 15 Jul 2014


%\cite{Fukushima:2008xe}
\bibitem{Fukushima:2008xe} 
  K.~Fukushima, D.~E.~Kharzeev and H.~J.~Warringa,
  %``The Chiral Magnetic Effect,''
  Phys.\ Rev.\ D {\bf 78}, 074033 (2008)
  [arXiv:0808.3382 [hep-ph]].
  %%CITATION = ARXIV:0808.3382;%%
  %426 citations counted in INSPIRE as of 15 Jul 2014

%\cite{Kharzeev:2013ffa}
\bibitem{Kharzeev:2013ffa} 
  D.~E.~Kharzeev,
  %``The Chiral Magnetic Effect and Anomaly-Induced Transport,''
  Prog.\ Part.\ Nucl.\ Phys.\  {\bf 75}, 133 (2014)
  [arXiv:1312.3348 [hep-ph]].
  %%CITATION = ARXIV:1312.3348;%
%\cite{Simonov:1997js}
\bibitem{Simonov:1997js}
  Yu.~A.~Simonov,
  %``Confinement,''
  Phys.\ Usp.\  {\bf 39} (1996) 313
   [Usp.\ Fiz.\ Nauk {\bf 166} (1996) 337]
  [hep-ph/9709344].
  %%CITATION = HEP-PH/9709344;%%
  %121 citations counted in INSPIRE as of 17 Mar 2013

%\cite{Leutwyler:1980ev}
\bibitem{Leutwyler:1980ev}
  H.~Leutwyler,
  %``Vacuum Fluctuations Surrounding Soft Gluon Fields,''
  Phys.\ Lett.\ B {\bf 96} (1980) 154;
  %%CITATION = PHLTA,B96,154;%%
  %69 citations counted in INSPIRE as of 22 Mar 2013
%\cite{Leutwyler:1980ma}
%\bibitem{Leutwyler:1980ma}
%  H.~Leutwyler,
  %``Constant Gauge Fields and their Quantum Fluctuations,''
  Nucl.\ Phys.\ B {\bf 179} (1981) 129.
  %%CITATION = NUPHA,B179,129;%%
  %135 citations counted in INSPIRE as of 17 Mar 2013


%\cite{Minkowski:1981ma} 
\bibitem{Minkowski:1981ma} 
  P.~Minkowski, 
  %``On The Ground State Expectation Value Of The Field Strength 
  %Bilinear In Gauge Theories And Constant Classical Fields,'' 
  Nucl.\ Phys.\ B {\bf 177} (1981) 203. 
  %%CITATION = NUPHA,B177,203;%% 
  %35 citations counted in INSPIRE as of 18 Mar 2013


%\cite{Flory:1983dx}
\bibitem{Flory:1983dx}
  C.~A.~Flory,
  %``A Selfdual Gauge Field, Its Quantum Fluctuations, And Interacting
  %Fermions,''
  Phys.\ Rev.\ D {\bf 28} (1983) 1425.
  %%CITATION = PHRVA,D28,1425;%%
  %27 citations counted in INSPIRE as of 17 Mar 2013

%\cite{vanBaal:1984ar}
\bibitem{vanBaal:1984ar}
  P.~van Baal,
  %``Su(n) Yang-mills Solutions With Constant Field Strength On
  %T**4,''
  Commun.\ Math.\ Phys.\  {\bf 94} (1984) 397.
  %%CITATION = CMPHA,94,397;%%
  %69 citations counted in INSPIRE as of 18 Mar 2013

%\cite{Efimov:1998hi}
\bibitem{Efimov:1998hi}
  G.~V.~Efimov, A.~C.~Kalloniatis and S.~N.~Nedelko,
  %``Confining properties of the homogeneous selfdual field and the
  %effective potential in SU(2) Yang-Mills theory,''
  Phys.\ Rev.\ D {\bf 59} (1998) 014026
  [hep-th/9806165].
  %%CITATION = HEP-TH/9806165;%%
  %15 citations counted in INSPIRE as of 17 Mar 2013

%%\cite{Ambjorn:1979xi}
\bibitem{Ambjorn:1979xi} 
  J.~Ambjorn and P.~Olesen,
  %``On the Formation of a Random Color Magnetic Quantum Liquid in QCD,''
  Nucl.\ Phys.\ B {\bf 170}, 60 (1980).
  %%CITATION = NUPHA,B170,60;%%
  %217 citations counted in INSPIRE as of 15 Jul 2014
  %13 citations counted in INSPIRE as of 15 Jul 2014

%\cite{Amundsen:1990nt}
\bibitem{Amundsen:1990nt} 
  P.~A.~Amundsen and M.~Schaden,
  %``Classical vacua in the field strength formulation of QCD,''
  Phys.\ Lett.\ B {\bf 252}, 265 (1990).
  %%CITATION = PHLTA,B252,265;%%
  %11 citations counted in INSPIRE as of 15 Jul 2014



%\cite{Liu:1999fc}
\bibitem{Liu:1999fc}
  H.~Liu and A.~A.~Tseytlin,
  %``D3-brane D instanton configuration and N=4 superYM theory in
  %constant selfdual background,''
  Nucl.\ Phys.\ B {\bf 553} (1999) 231
  [hep-th/9903091].
  %%CITATION = HEP-TH/9903091;%%
  %64 citations counted in INSPIRE as of 18 Mar 2013

%\cite{Kehagias:1999iy}
\bibitem{Kehagias:1999iy}
  A.~Kehagias and K.~Sfetsos,
  %``On asymptotic freedom and confinement from type IIB
  %supergravity,''
  Phys.\ Lett.\ B {\bf 456} (1999) 22
  [hep-th/9903109].
  %%CITATION = HEP-TH/9903109;%%
  %57 citations counted in INSPIRE as of 18 Mar 2013

%\cite{Ghoroku:2004sp}
\bibitem{Ghoroku:2004sp}
 K.~Ghoroku and M.~Yahiro,
  %``Chiral symmetry breaking driven by dilaton,''
  Phys.\ Lett.\ B {\bf 604} (2004) 235
  [hep-th/0408040].
  %%CITATION = HEP-TH/0408040;%%
  %76 citations counted in INSPIRE as of 18 Mar 2013

%\cite{Ghoroku:2005tf}
\bibitem{Ghoroku:2005tf}
  K.~Ghoroku, T.~Sakaguchi, N.~Uekusa and M.~Yahiro,
  %``Flavor quark at high temperature from a holographic model,''
  Phys.\ Rev.\ D {\bf 71} (2005) 106002
  [hep-th/0502088].
  %%CITATION = HEP-TH/0502088;%%
  %64 citations counted in INSPIRE as of 19 Mar 2013

%\cite{Brevik:2005fs}
\bibitem{Brevik:2005fs}
  I.~H.~Brevik, K.~Ghoroku, A.~Nakamura,
  %``Meson mass and confinement force driven by dilaton,''
  Int.\ J.\ Mod.\ Phys.\ D {\bf 15} (2006) 57
  [hep-th/0505057].
  %%CITATION = HEP-TH/0505057;%%
  %30 citations counted in INSPIRE as of 04 Apr 2013

%\cite{Karch:2002sh}
%\bibitem{Karch:2002sh}
%  A.~Karch and E.~Katz,
  %``Adding flavor to AdS / CFT,''
%  JHEP {\bf 0206} (2002) 043
%  [hep-th/0205236].
  %%CITATION = HEP-TH/0205236;%%
  %595 citations counted in INSPIRE as of 19 Mar 2013

%\cite{Ghoroku:2006af}
\bibitem{Ghoroku:2006af}
  K.~Ghoroku, M.~Ishihara and A.~Nakamura,
  %``Gauge theory in de Sitter space-time from a holographic model,''
  Phys.\ Rev.\ D {\bf 74} (2006) 124020
  [hep-th/0609152].
  %%CITATION = HEP-TH/0609152;%%
  %7 citations counted in INSPIRE as of 20 Mar 2013

%\cite{Erdmenger:2007vj}
\bibitem{Erdmenger:2007vj}
  J.~Erdmenger, K.~Ghoroku and I.~Kirsch,
  %``Holographic heavy-light mesons from non-Abelian DBI,''
  JHEP {\bf 0709} (2007) 111
  [arXiv:0706.3978 [hep-th]].
  %%CITATION = ARXIV:0706.3978;%%
  %11 citations counted in INSPIRE as of 20 Mar 2013
%\cite{Erdmenger:2011sz}
\bibitem{Erdmenger:2011sz}
  J.~Erdmenger, A.~Gorsky, P.~N.~Kopnin, A.~Krikun and A.~V.~Zayakin,
  %``Low-Energy Theorems from Holography,''
  JHEP {\bf 1103} (2011) 044
  [arXiv:1101.1586 [hep-th]].
  %%CITATION = ARXIV:1101.1586;%%
  %6 citations counted in INSPIRE as of 20 Mar 2013
%\cite{Ghoroku:2008tg}

\bibitem{Ghoroku:2008tg}
  K.~Ghoroku and M.~Ishihara,
  %``Baryons with D-5-brane vertex and k-quark states,''
  Phys.\ Rev.\ D {\bf 77} (2008) 086003
  [arXiv:0801.4216 [hep-th]].
  %%CITATION = ARXIV:0801.4216;%%
  %18 citations counted in INSPIRE as of 20 Mar 2013

%\cite{Ghoroku:2008na}
\bibitem{Ghoroku:2008na}
  K.~Ghoroku, M.~Ishihara, A.~Nakamura and F.~Toyoda,
  %``Multi-quark baryons and color screening at finite temperature,''
  Phys.\ Rev.\ D {\bf 79} (2009) 066009
  [arXiv:0806.0195 [hep-th]].
  %%CITATION = ARXIV:0806.0195;%%
  %12 citations counted in INSPIRE as of 20 Mar 2013

%\cite{Sin:2009yu}
\bibitem{Sin:2009yu}
  S.~-J.~Sin, S.~Yang and Y.~Zhou,
  %``Comments on Baryon Melting in Quark Gluon Plasma with Gluon Condensation,''
  JHEP {\bf 0911} (2009) 001
  [arXiv:0907.1732 [hep-th]].
  %%CITATION = ARXIV:0907.1732;%%
  %4 citations counted in INSPIRE as of 19 Mar 2013

%\cite{Gwak:2012ht}
\bibitem{Gwak:2012ht}
  B.~Gwak, M.~Kim, B.~-H.~Lee, Y.~Seo and S.~-J.~Sin,
  %``Holographic D Instanton Liquid and chiral transition,''
  Phys.\ Rev.\ D {\bf 86} (2012) 026010
  [arXiv:1203.4883 [hep-th]].
  %%CITATION = ARXIV:1203.4883;%%
  %2 citations counted in INSPIRE as of 18 Mar 2013

%\cite{Sin:2009dk}
\bibitem{Sin:2009dk}
  S.~-J.~Sin and Y.~Zhou,
  %``Holographic melting of Heavy Baryons in Plasma with Gluon Condensation,''
  JHEP {\bf 0905} (2009) 044
  [arXiv:0904.4249 [hep-th]].
  %%CITATION = ARXIV:0904.4249;%%
  %5 citations counted in INSPIRE as of 20 Mar 2013

%\cite{Witten:1998zw}
\bibitem{Witten:1998zw}
  E.~Witten,
  %``Anti-de Sitter space, thermal phase transition, and confinement
  %in gauge theories,''
  Adv.\ Theor.\ Math.\ Phys.\  {\bf 2} (1998) 505
  [hep-th/9803131].
  %%CITATION = HEP-TH/9803131;%%
  %1806 citations counted in INSPIRE as of 19 Mar 2013

%\cite{Sakai:2004cn}
\bibitem{Sakai:2004cn}
  T.~Sakai and S.~Sugimoto,
  %``Low energy hadron physics in holographic QCD,''
  Prog.\ Theor.\ Phys.\  {\bf 113} (2005) 843
  [hep-th/0412141].
  %%CITATION = HEP-TH/0412141;%%
  %694 citations counted in INSPIRE as of 20 Mar 2013

%\cite{Sakai:2005yt}
\bibitem{Sakai:2005yt}
  T.~Sakai and S.~Sugimoto,
  %%``More on a holographic dual of QCD,''
  Prog.\ Theor.\ Phys.\  {\bf 114} (2005) 1083
  [hep-th/0507073].
  %%CITATION = HEP-TH/0507073;%%
  %%414 citations counted in INSPIRE as of 20 Mar 2013


%\cite{Barbon:1999zp}
\bibitem{Barbon:1999zp}
  J.~L.~F.~Barb\'on and A.~Pasquinucci,
  %``Aspects of instanton dynamics in AdS / CFT duality,''
  Phys.\ Lett.\ B {\bf 458} (1999) 288
  [hep-th/9904190].
  %%CITATION = HEP-TH/9904190;%%
  %8 citations counted in INSPIRE as of 20 Mar 2013
%\cite{Suzuki:2000sv}
\bibitem{Suzuki:2000sv}
  K.~Suzuki,
  %``D0 - D4 system and QCD(3+1),''
  Phys.\ Rev.\ D {\bf 63} (2001) 084011
  [hep-th/0001057].
  %%CITATION = HEP-TH/0001057;%%
  %2 citations counted in INSPIRE as of 20 Mar 2013

%\cite{Wu:2013zxa}
\bibitem{Wu:2013zxa} 
  C.~Wu, Z.~Xiao and D.~Zhou,
  %``Sakai-Sugimoto model in D0-D4 background,''
  Phys.\ Rev.\ D {\bf 88}, no. 2, 026016 (2013)
  [arXiv:1304.2111 [hep-th]].
  %%CITATION = ARXIV:1304.2111;%%
  %1 citations counted in INSPIRE as of 15 Jul 2014


%\cite{Witten:1979vv}
%\bibitem{Witten:1979vv}
%  E.~Witten,
  %``Current Algebra Theorems for the U(1) Goldstone Boson,''
%  Nucl.\ Phys.\ B {\bf 156} (1979) 269.
  %%CITATION = NUPHA,B156,269;%%
  %1097 citations counted in INSPIRE as of 19 May 2013

%\cite{Zahed:1986qz}
%\bibitem{Zahed:1986qz}
%  I.~Zahed, G.~E.~Brown and ,
  %``The Skyrme Model,''
%  Phys.\ Rept.\  {\bf 142} (1986) 1.
  %%CITATION = PRPLC,142,1;%%
  %493 citations counted in INSPIRE as of 25 Mar 2013

%\cite{Gabadadze:2004jq}
%\bibitem{Gabadadze:2004jq}
%  G.~Gabadadze, A.~Iglesias and ,
  %``On theta dependence of glueballs from AdS / CFT,''
%  Phys.\ Lett.\ B {\bf 609} (2005) 167
%  [hep-th/0411278].
  %%CITATION = HEP-TH/0411278;%%
  %8 citations counted in INSPIRE as of 30 Mar 2013

%\cite{Bergshoeff:2005zf}
\bibitem{Bergshoeff:2005zf}
  E.~Bergshoeff, A.~Collinucci, A.~Ploegh, S.~Vandoren and T.~Van Riet,
  %``Non-extremal D-instantons and the AdS/CFT correspondence,''
  JHEP {\bf 0601} (2006) 061
  [hep-th/0510048].
  %%CITATION = HEP-TH/0510048;%%
  %14 citations counted in INSPIRE as of 04 Apr 2013

%\cite{Nawa:2006gv}
\bibitem{Nawa:2006gv} 
  K.~Nawa, H.~Suganuma and T.~Kojo,
  %``Baryons in holographic QCD,''
  Phys.\ Rev.\ D {\bf 75}, 086003 (2007)
  [hep-th/0612187].
  %%CITATION = HEP-TH/0612187;%%
  %84 citations counted in INSPIRE as of 15 Jul 2014 
%\cite{Seki:2013nta}

%\cite{Hata:2007mb}
\bibitem{Hata:2007mb}
  H.~Hata, T.~Sakai, S.~Sugimoto and S.~Yamato,
  %``Baryons from instantons in holographic QCD,''
  Prog.\ Theor.\ Phys.\  {\bf 117} (2007) 1157
  [hep-th/0701280 [HEP-TH]].
  %%CITATION = HEP-TH/0701280;%%
  %160 citations counted in INSPIRE as of 20 Mar 2013

%\cite{Hashimoto:2008zw}
\bibitem{Hashimoto:2008zw}
  K.~Hashimoto, T.~Sakai and S.~Sugimoto,
  %``Holographic Baryons: Static Properties and Form Factors from
  %Gauge/String Duality,''
  Prog.\ Theor.\ Phys.\  {\bf 120} (2008) 1093
  [arXiv:0806.3122 [hep-th]].
  %%CITATION = ARXIV:0806.3122;%%
  %92 citations counted in INSPIRE as of 20 Mar 2013

%\cite{Hashimoto:2009ys}
\bibitem{Hashimoto:2009ys}
  K.~Hashimoto, T.~Sakai and S.~Sugimoto,
  %``Nuclear Force from String Theory,''
  Prog.\ Theor.\ Phys.\  {\bf 122} (2009) 427
  [arXiv:0901.4449 [hep-th]].
  %%CITATION = ARXIV:0901.4449;%%
  %34 citations counted in INSPIRE as of 20 Mar 2013


%\cite{Kaplunovsky:2010eh}
\bibitem{Kaplunovsky:2010eh}
  V.~Kaplunovsky and J.~Sonnenschein,
  %``Searching for an Attractive Force in Holographic Nuclear Physics,''
  JHEP {\bf 1105} (2011) 058
  [arXiv:1003.2621 [hep-th]].
  %%CITATION = ARXIV:1003.2621;%%
  %10 citations counted in INSPIRE as of 20 Mar 2013

\bibitem{Seki:2013nta} 
  S.~Seki and S.~-J.~Sin,
  %``A New Model of Holographic QCD and Chiral Condensate in Dense Matter,''
  JHEP {\bf 1310}, 223 (2013)
  [arXiv:1304.7097 [hep-th]].
  %%CITATION = ARXIV:1304.7097;%%
\end{thebibliography}
\end{document}